# The Evolution of X-ray Clusters in a Cold plus Hot Dark Matter Universe

Greg L. Bryan[1,2], Anatoly Klypin[3], Chris Loken[3]
Michael L. Norman[1,2], Jack O. Burns[3]

[1] *National Center for Supercomputing Applications*
*5600 Beckman Institute, Drawer 25*
*University of Illinois at Urbana-Champaign*
*405 North Mathews Avenue, Urbana, IL 61801*

[2] *Department of Astronomy*
*University of Illinois at Urbana-Champaign*

[3] *Department of Astronomy*
*New Mexico State University*
*Las Cruces, NM 88003-8001*

## ABSTRACT

We present the first self-consistently computed results on the evolution of X-ray properties of galaxy clusters in a Cold + Hot Dark Matter (CHDM) model. We have performed a hydrodynamic plus N-body simulation for the COBE-compatible CHDM model with standard mass components: $\Omega_{hot} = 0.3$, $\Omega_{cold} = 0.6$ and $\Omega_{baryon} = 0.1$ ($h = 0.5$). In contrast with the CDM model, which fails to reproduce the observed temperature distribution function $dN/dT$ (Bryan et al. 1994b), the CHDM model fits the observational $dN/dT$ quite well. Our results on X-ray luminosity are less firm but even more intriguing. We find that the resulting X-ray luminosity functions at redshifts $z = 0.0, 0.2, 0.4, 0.7$ are well fit by observations, where they overlap. The fact that both temperatures and luminosities provide a reasonable fit to the available observational data indicates that, unless we are missing some essential physics, there is neither room nor need for a large fraction of gas in rich clusters: 10% (or less) in baryons is sufficient to explain their X-ray properties. We also see a tight correlation between X-ray luminosity and gas temperature.

*Subject headings:* Cosmology: large-scale structure of the Universe — hydrodynamics — X-ray: general

## 1. INTRODUCTION

X-ray observations of clusters of galaxies provide an important cosmological probe which can, for example, be used to estimate the amplitude and shape of the fluctuation spectrum (Henry & Arnaud 1991), the mean density of the universe (Richstone et al. 1992) and the fraction of mass in baryons (White 1992 and White et al. 1993). X-ray clusters have a reduced likelihood of confusion with the background and of chance projection effects from smaller clusters, an occurrence which can plague optical studies (Frenk et al. 1990; Dekel et al. 1989). It is also easier to self-consistently model the X-ray emitting gas which is thought to dominate over the mass in galaxies (White 1992) than to follow the evolution of the much smaller and denser galaxies.

The CHDM scenario has scored a number of successes in explaining observations: the galaxy correlation function, galaxy pairwise velocities, bulk velocities (Klypin et al. 1993 = KHPR), the number and virial properties of groups (Nolthenius et al. 1994), the cluster-cluster correlation function (Holtzman & Primack 1993; Klypin & Rhee 1994). Perhaps its severest challenge comes from the large number of damped Lyman-$\alpha$ systems observed at moderate to high redshifts (Lanzetta 1993, Lanzetta et al. 1993). Mo & Miralda-Escudé (1994) and Kauffmann & Charlot (1994) have claimed that the model predicts too few high redshift damped Lyman-$\alpha$ systems ($z \approx 3$). The situation depends on the identification of the observational data at these redshifts: large galaxies (Briggs et al. 1989) or dwarfs (Hunstead, Pettini, & Fletcher 1990). Klypin et al. (1994) argue that the existence of many large galaxies at z=3 (comparable with the present-day number and masses of galaxies) would not leave room for much needed subsequent evolution. As an alternative, they propose a variant of the CHDM model ($\Omega_{cold}/\Omega_{hot}/\Omega_{baryon} = 0.675/0.25/0.075$), which has basically the same properties on large scales and predicts more high redshift objects as compared with the standard CHDM.

In this paper we test the standard CHDM model at cluster lengths for both the dark matter and baryonic components. Predictions of cluster parameters for the model have been made by a number of groups using a variety of methods. The correlation function was estimated by both analytical (Holtzman & Primack 1993) and numerical methods (Jing & Fang 1994; Jing et al. 1993; Klypin & Rhee 1994; Cen & Ostriker 1994). Predictions for X-ray emission were made by Bartlett & Silk (1993) using the Press-Schecter approximation and by Klypin & Rhee (1994) using N-body simulations and assuming the cluster gas temperature was proportional to the square of the cold component's velocity dispersion.

To our knowledge, this paper presents the first self-consistent hydrodynamic plus N-body treatment of X-ray clusters for the CHDM model. We stress the importance of such simulations. Clusters in the model (and probably in the real universe) form at relatively low redshifts, therefore one should expect that the results of recent violent creation, mergers, and accretion will be of importance.

## 2. SIMULATIONS

The numerical method used to simulate the gas has been described in detail elsewhere (Bryan et al. 1994a), as has the general scheme for generating initial conditions (KHPR), so we restrict ourselves to a brief overview. The cold and hot components were followed with a standard particle-mesh (PM) simulation using a $256^3$ grid with $128^3$ cold particles and $2 \times 128^3$ hot particles in order to better sample the neutrino phase space. The gas was represented with a $256^3$ grid and evolved with the higher-order accurate piecewise-parabolic method. This is an Eulerian scheme that features third order advection with multi-dimensional shock capturing and has been modified and extensively tested for cosmological applications. We did not include any radiative effects as this will have little effect on the hot cluster gas which has a cooling timescale on order of the Hubble time.

The initial power spectrum and growth rate came from analytic fits given in KHPR ("note added in proof"), with the hot particles being set up in pairs. Each particle in a pair had the same initial position and perturbation-driven velocity but were given equal but opposite thermal velocities drawn from relativistic Fermi-Dirac statistics. We used a box size of 85 $h^{-1}$ Mpc, with $\Omega_{hot} = 0.3$, $\Omega_{cold} = 0.6$ and $\Omega_{baryon} = 0.1$. We write the Hubble constant $H = 100h$ km s$^{-1}$ Mpc$^{-1}$ and use $h = 0.5$ throughout. A normalization which produces a 15.5 $\mu$K rms quadrapole of the microwave background anisotropy was chosen. This corresponds to a bias parameter of $b = 1.65$, very close to that given by COBE. We began the simulation at a redshift $z = 15$.

Clusters are identified through their X-ray signature: the thermal Bremsstrahlung emission is computed for each cell in the energy band of interest. We assume primordial abundances (76% H and 24% He by mass) with both hydrogen and helium fully ionized (Spitzer 1978). We neglect lines; however, tests with a full ionization equilibrium code (Raymond & Smith 1977) indicate that we are underestimating the luminosity by only about 20–40%. The free-free Gaunt factor is approximated with numerical fits given by Kellogg, Baldwin & Koch (1975).

Cells are tagged as potential cluster centers provided they have a luminosity greater than $10^{38}$ erg s$^{-1}$ and are local maxima compared to all 26 neighbouring cells. We then attempt to estimate the cluster center by finding the center-of-mass within a $0.5h^{-1}$ Mpc sphere. This is done by dividing the sphere into a new grid of zones in spherical coordinates, equally spaced in both the azimuthal direction and the cosine of the polar angle, but logarithmically spaced in radius. The density in each cell is found by a linear interpolation from the nearest original grid points. After the center-of-mass is found, the procedure is repeated at the new position until convergence occurs. Once the list is constructed, clusters whose centers are separated by less than a comoving distance of 2 $h^{-1}$ Mpc are merged. Cluster characteristics such as mass and luminosity are then computed by a similar technique, integrating over a sphere with a comoving radius of $1h^{-1}$ Mpc.

We have also tested a number of other schemes for identifying cluster centers and computing their properties, including the procedure described in Kang et al. (1994). We find that there is little sensitivity to methodology although the Kang et al. technique for computing the luminosity-weighted temperature appears to slightly overestimate cluster temperatures as compared to other schemes. We also note that the results are largely unchanged if we increase the cutoff radii, indicating that we are capturing the majority of the emission.

In order to improve our ability to compare with observations, we have computed a large PM CHDM realization with $768^3$ zones and $256^3$ particles in a box 255 $h^{-1}$ Mpc on a side, giving us the same resolution as in the hydrodynamic simulation. Although we do not have thermodynamic information, we can fix the relation between the one-dimensional dark matter velocity dispersion and temperature from the smaller gas dynamical simulation. We adopt the relation $T(\text{keV}) = (v/v_*)^2$ where $v$ is the cold particle one-dimensional velocity dispersion in km/s, computed in spheres of 1.5 $h^{-1}$ Mpc and $v_* = 410$ km/s. The value of $v_*$ seems to be slightly scale dependent, therefore we fix this normalization with the largest clusters in our 85 $h^{-1}$ Mpc box. Because there are a limited number of such clusters, the uncertainty in the quoted $v_*$ is relatively large. Similarly, one can obtain a 2-10 keV luminosity-velocity relation. We use the following fit to the data:

$$L = 5 \times 10^{44} \left(\frac{v}{1000 \text{ km/s}}\right)^7 \text{ erg/s}. \tag{1}$$

This is compatible with the adopted $T$–$v$ relation combined with the observed luminosity temperature relation, $L \propto T^{\sim 3.5}$, and provides a good fit to the brighter ($L_{2-10} > 10^{42}$ erg/s) simulated clusters.

## 3. RESULTS AND DISCUSSION

### 3.1. Luminosity and Temperature Distribution Functions

In Figure 1, we show the 2-10 keV luminosity-weighted temperature distribution function for gas in clusters at $z = 0$ and compare it to data from the volume-limited sample of Edge et al. (1990). The fit is quite good. There are almost no high temperature clusters ($T \sim 10$ keV), however these clusters correspond to rare events in the initial density field and we believe that their absence is simply due to the limited volume of our simulated box (which is much smaller than the observational survey). Also shown are the results from the larger PM simulation with the normalization described above. Within uncertainties, they agree well with observations over a wide range of length scales. The cluster temperature, which we find to be nearly constant over the entire cluster, is our most reliable result, as the numerical scheme conserves energy and is adept at correctly modeling strong shocks.

The 2-10 keV luminosity function at the present epoch is plotted in Figure 2a, along with data from Edge et al. (1990). Again, our box is too small to contain the large, rare clusters; however, in the region of overlap, the fit is reasonably good.

In order to investigate the evolution of the clusters, we compute the luminosity function in the 0.3-3.5 keV range (observer's frame) at $z = 0.2$ (Figure 2b) and compare it to the *Einstein* Extended Medium Sensitivity Survey (Henry et al. 1992 and Henry 1992). The authors present evidence for the evolution of the luminosity function in three redshift shells; unfortunately, the strongest changes occur for the brightest clusters, which are outside of our luminosity range.

In Figure 2c we plot the luminosity function in the ROSAT bandpass (0.5–2.5 keV) at a redshift of $z = 0.5$. Also plotted is the $z = 0.41$ luminosity function from Bower et al. (1994). We also note that pointed observations of likely cluster candidates (Nichol et al. 1994 and Castander et al. 1993) have produced a lower limit for the number density of X-ray clusters of $1.2 \times 10^{-7}$ Mpc$^{-3}$ for clusters with a luminosity around $10^{44}$ erg/s. Our luminosity function for the same epoch is consistent with this number which represents a substantial evolutionary effect. This is reinforced by Figure 2d which shows the strong negative evolution with increasing redshift predicted in CHDM. This provides a strong check on the model: COBE normalized CDM, for example, predicts weak evolution in the opposite sense (Bryan et al. 1994b; Kang et al. 1994). The cluster luminosity is not as well constrained as the temperature and it is possible that we have substantially underestimated the cluster luminosity (see below).

The relation between a cluster's temperature and its luminosity should be less sensitive to the power spectrum and normalization. We show this relation in Figure 3 along with observational points from David, et al. (1993). The slopes match quite closely; however, due to the uncertainties in our cluster luminosities, it is not clear how well the simulated slope is determined. In order for the curves to align, it must be assumed that we are underestimating the luminosities by a factor of two, but this is well within our systematic uncertainties for computed cluster luminosities.

### 3.2. Discussion of Errors

With a cell size of 330 $h^{-1}$ kpc, we obviously cannot fully resolve a cluster. In order to gauge the extent of the error committed, we have performed some tests using smaller physical boxes with higher resolution and find that cluster temperatures are very well constrained, but central cluster densities increase as the resolution increases. Since free-free emission is proportional to the square of the density, this means that formally, we can only calculate a lower bound to a cluster's true luminosity. We are also performing a more complete resolution study (Anninos & Norman 1994) which seems to imply that without other physical processes (such as star formation), we do not produce resolvable cores (at least to a resolution level of $50 - 100$ $h^{-1}$ kpc).

Real clusters, however, do have identifiable cores, therefore we must ask how our results compare to clusters as seen by X-ray telescopes. These seem to follow a profile of the form (Jones & Forman 1984)

$$\rho = \rho_0[1 + (r^2/R_c^2)]^{-3\beta/2}, \qquad (2)$$

with $\beta \sim 0.8$ and $R_c \sim 150$ $h^{-1}$ kpc. This assumes that the clusters are roughly isothermal,

a conclusion in reasonable agreement with the simulation. In order to compare this typical observational cluster, we fit the same profile to our simulated clusters, obtaining values of $\beta \sim 1.4$ and $R_c \sim 0.7\ h^{-1}$ Mpc. These parameters are quite different from most observed clusters, but how much does this affect the resulting luminosity? To compare, we must specify the remaining free parameter: $\rho_0$. This is done by demanding that both clusters (simulated and observed) have the same mass within a 1 $h^{-1}$ Mpc radius. Despite the extremely disparate core sizes, the real cluster's luminosity is only about three times higher.

Besides resolution, the absence of a number of physical effects could change the calculated luminosity. These include: line emission from metals, although the omission of this effect is probably limited to an increase of the luminosity by 20–40%; radiative cooling, which would enhance the flux in denser, cooler regions; and galaxy formation, which would decrease the available baryons and hence the luminosity of the cluster.

## 4. CONCLUSIONS

By performing hydrodynamic plus N-body simulations of the CHDM model, we have removed a layer of assumptions between theory and observations. The simulation directly provides the thermodynamic state of the hot gas in clusters, allowing us to test the model against easily understood observations. We find that the CHDM model is consistent with X-ray observations of cluster-sized perturbations in the available redshift range under a few well-motivated assumptions.

These results, however, are quite sensitive to our choice of $\Omega_{baryon}$ since the luminosity depends on the square of the baryon density. For our choice of h = 0.5, light element nucleosynthesis predicts $\Omega_{baryon} = 0.05$ (Walker et al. 1990). If this figure were adopted it would imply a global decrease in the computed luminosity by about a factor of four. These large factors may in fact be compatible with the model due to the possibility that we are substantially underestimating cluster luminosities (because of limited resolution). We are not, on the other hand, overestimating luminosities, therefore a higher $\Omega_{baryon} (> 0.1)$, as might be indicated by cluster virial estimates (White 1992) is incompatible with the model as it would push the predicted luminosity function above observations.

As a caution, we observe that it is not yet possible to have the large dynamic range required to both accurately compute the internal physics of an individual cluster and to simulate a sufficient region to include the brightest clusters, although the results presented here show that we have made significant progress towards this goal. Simulations with higher resolution and larger boxes are in progress (Bryan & Norman 1994).

Finally, we note that it would be helpful to have an observational luminosity (and temperature) function that extended to lower luminosities, to circumvent the difficulties of simulating large volumes of space. Thus we stress the need for X-ray observations of poor clusters and groups.

4– 7 –

We are happy to acknowledge support from NSF grant ASC93-18185, and NASA Long Term Space Astrophysics Program grant NAGW-3152. The simulations were performed on the Convex C3880 supercomputer at the National Center for Supercomputing Applications.## REFERENCES

Anninos, P., Norman, M.L. 1994, in preparation

Bartlett, J.G, Silk, J 1993, ApJ, 407, L45

Bower, R.G., Böhringer, Briel, U.G., Ellis, R.S., Castander, F.J., & Couch, W.J. 1994, MNRAS, in press

Briggs, F.H., Wolfe, A.M., Liszt, H.S., Davis, H.M., Turner, K.L. 1989, ApJ, 341, 650

Bryan, G.L., & Norman, M.L., 1994, in preparation

Bryan, G.L., Norman, M.L., Stone, J.M., Cen, R.Y., & Ostriker, J.P. 1994a, Computer Physics Communications, in press

Bryan, G.L., Cen, R., Norman, M.L., Ostriker, J.P., & Stone, J.M. 1994b, ApJ, 428, 405

Cen, R., & Ostriker, J.P. 1994, ApJ, in press

Castander, F.J., Ellis, R., S., Frenk, C.S., Dressler, A. & Gunn, J.E. 1993, ApJ 424, L79

David, L.P., Slyz, A., Jones, C., Forman, W., & Vrtilek, S.D. 1993, ApJ, 412, 479

Dekel, A., Blumenthal, G. Primack, J.R., & Oliver, S. 1989, ApJ, 338, L5

Edge, A.C., Stewart, G.C., Fabina, A.C., Arnaud, K.A. 1990, MNRAS, 245, 559

Frenk, C.S., White, S.D.M., Efstathiou, G., & Davis, M. 1990, ApJ, 351, 10

Henry, J.P. 1992, in "Clusters and Superclusters of Galaxies" (Kluwer Publishers: Dordrecht), p.311, ed. A.C. Fabian

Henry, J.P., & Arnaud, K.A. 1991, ApJ, 372, 410

Henry, J.P., Gioia, I.M, Maccacaro, T., Morris, S.L., Stocke, J.T., & Wolter, A. 1992, ApJ, 386, 408

Holtzman, J.A., Primack, J.R. 1993, ApJ, 405, 428

Hunstead, R.W., Pettini, M., & Fletcher, A.B. 1990, ApJ, 356, 23

Jing, Y.P. & Fang, L.Z. 1994, ApJ, in press

Jones, C., & Forman, W. 1984, ApJ, 276, 38

Jing, Y.P., Mo, H.J., Borner, G., Fang, L.Z. 1993, ApJ, 411, 450

Kang, H., Cen, R., Ostriker, J.P., & Ryu, D. 1994, ApJ, 428, 1

Kauffmann, G. & Charlot, S. 1994, preprint

Kellogg, E., Baldwin, J.R., & Koch, D. 1975, ApJ, 199, 299

Klypin, A., Holtzman, J., Primack, J.R., & Regös, E. 1993, ApJ, 416, 1 (KHPR)

Klypin, A., & Rhee, G. 1994, ApJ, 428, 00

Klypin, A., Borgani, S., Holtzman, J., Primack, J. 1994, ApJ, submitted (astro-ph 9405003)

Lanzetta, K.M. 1993, PASP, 105, 1063

Lanzetta, K.M., Wolfe, A.M., Turnshek, D.A. 1993, ApJ, submitted

Mo, H.J. & Miralda-Escudé, J. 1994, ApJL, submitted (astro-ph preprint 9402014)

Nichol, R.C., Ulmer, M.P., Kron R.G., Wirth, G.D., & Koo, D.C. 1994, preprint

Nolthenius, R., Klypin, A., Primack, J. 1994, ApJ, 422, L45

Raymond, J.C., and Smith, B.W. 1977, ApJS, 35, 419.

Richstone, D., Loeb, A., & Turner, E.L. 1992, ApJ, 413, 492

Spitzer, L. Jr. 1978, Physical Processes in the Interstellar Medium (New York: Wiley)

Walker, T.P., Steigman, G., Schramm, D.N., Olive, K.A., & Kang, H.S. 1990, ApJ, 376, 51

White, S.D.M. 1992 in "Clusters and Superclusters of Galaxies" (Kluwer Publishers: Dordrecht), p.17, ed. A.C. Fabian

White, S.D.M., Navarro, J.N., Evrard, A.E., & Frenk, C.S. 1993, Nature, 366, 429

---

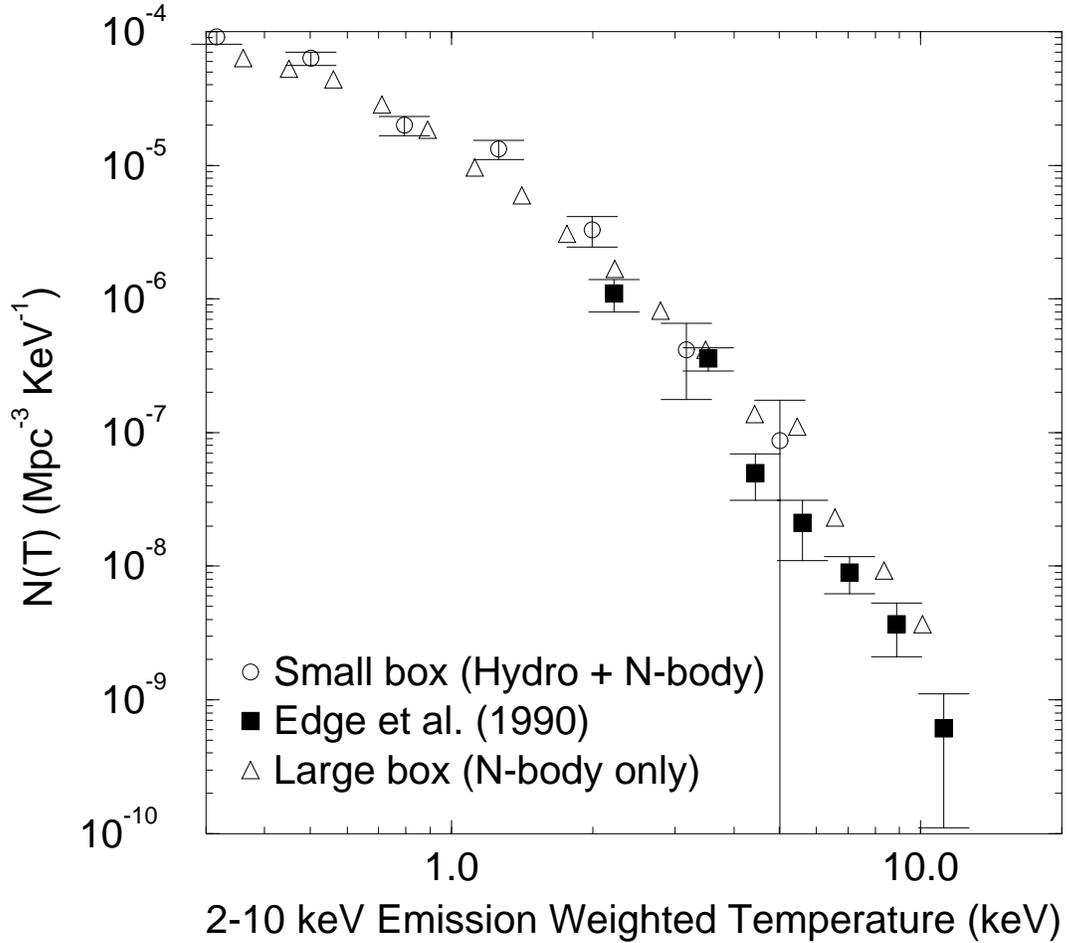

Fig. 1.— The 2–10 keV luminosity-weighted temperature distribution function at $z = 0$ with $\Omega_{baryon} = 0.1$ and $h = 0.5$. The open circles are from the 85 $h^{-1}$ Mpc box computed with our full hydrodynamics plus N-body code. The triangles are from a particle-mesh simulation with a 255 $h^{-1}$ Mpc box normalized to the results from the smaller box (see text). The open squares are points from the volume-limited sample of the Edge et al. (1990) data.

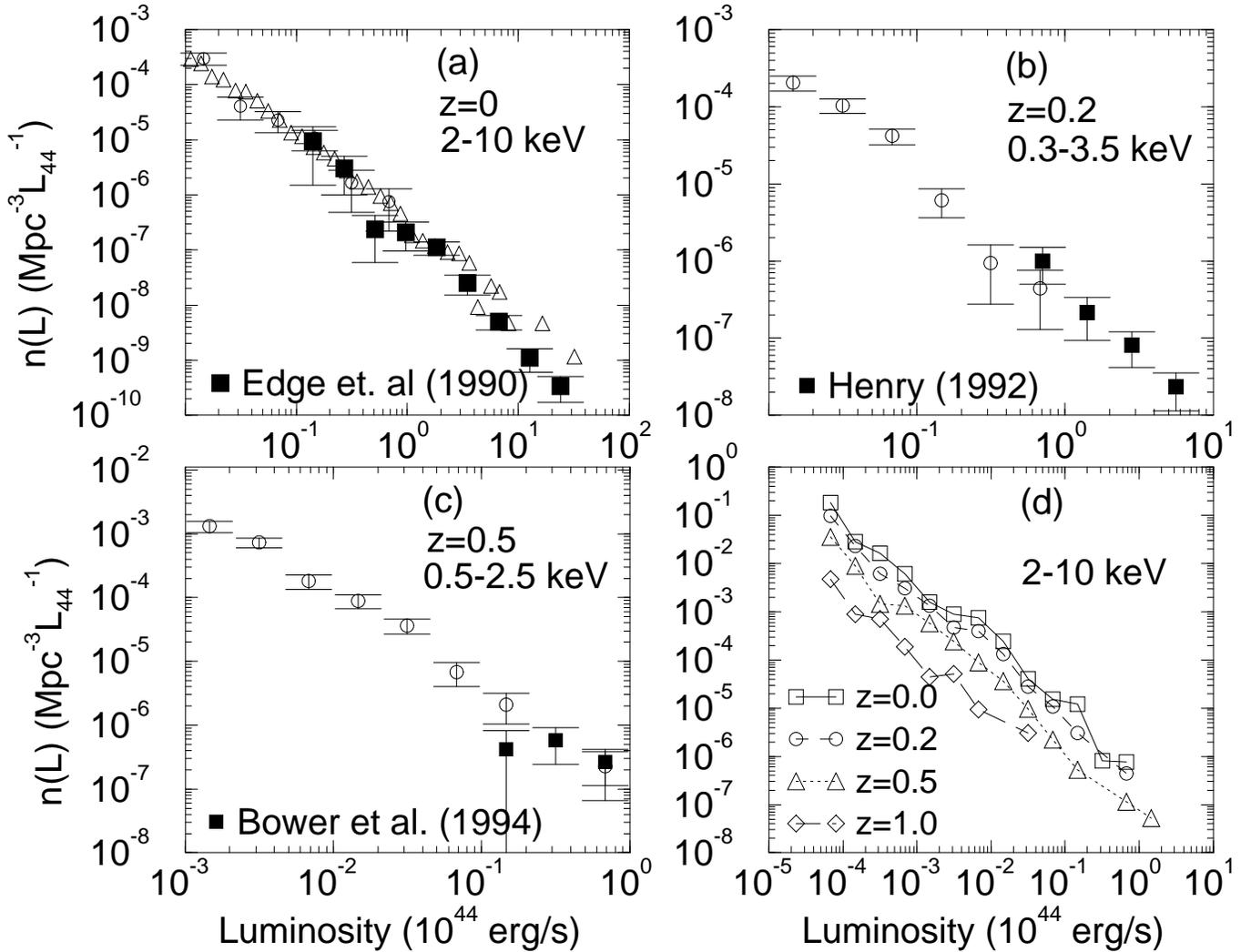

Fig. 2.— Luminosity functions at a variety of redshifts with $\Omega_{baryon} = 0.1$ and $h = 0.5$. In (a-c) open symbols are as in Fig. 1, while filled squares are observational data. We show luminosity functions in the (cluster rest frame's) energy range (a), 2–10 keV ($z = 0$), (b) 0.3–3.5 keV keV ($z = 0.2$), (c) 0.5–2.5 keV ($z = 0.5$). In (d) we show the predicted evolution of the 2–10 keV luminosity function with redshift.

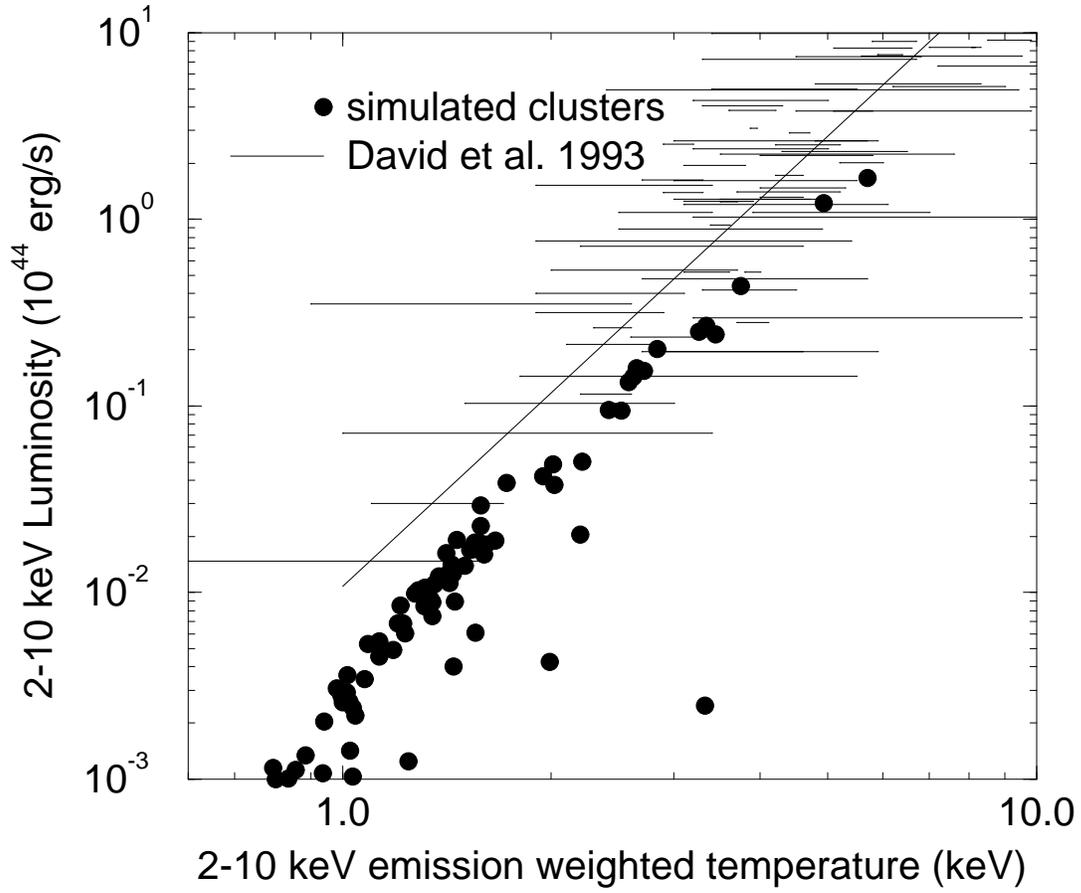

Fig. 3.— The luminosity-temperature relation at $z = 0$. The filled circles are simulated clusters while the temperature of clusters from the David et al. (1993) sample are plotted to show the uncertainty in the observed L-T relation. The solid line is the best fit power law to the observed distribution.